\documentclass[10pt,twoside]{article}
\usepackage{graphicx}
\usepackage{latexsym}
\usepackage{amsfonts}
\usepackage{amssymb}
\usepackage{amsmath}

\makeindex

\begin{document}

\title{Braneworld Cosmology}

\markboth{Jiro Soda and Sugumi Kanno}{Braneworld Cosmology}

\author{Jiro Soda and Sugumi Kanno
\\[5mm]
\it Department of Physics, Kyoto University, Kyoto 606-8501, Japan\\
\it e-mail: jiro@tap.scphys.kyoto-u.ac.jp, sugumi@tap.scphys.kyoto-u.ac.jp\\
}

\date{}
\maketitle

\thispagestyle{empty}

\begin{abstract}
\noindent
We present a general formalism for studying an inflationary scenario  
in the two-brane system. We explain the gradient expansion method
 and obtain the 4-dimensional effective action. Based on this
 effective action, we give a basic equations for the background
 homogeneous universe  and the cosmological perturbations.
 As an application, we propose a born-again braneworld scenario. 
 
\end{abstract}

\section{Introduction}

It is widely believed that the initial singularity problem in the conventional
 standard cosmology can be resolved by the quantum theory of gravity.
 It is also known that the most promising candidate for the quantum theory 
 of gravity is the superstring theory. Remarkably, the superstring theory 
 precisely predicts the existence of extra-dimensions.  
  As our universe can be recognized as  the 4-dimensional 
  Lorentzian manifold by experience, the necessity for reconciling 
  the prediction of the superstring theory with our experience
 is apparent. 
 
 The solution to this problem was proposed by Kaluza and Klein 
 a long time ago. Their idea is simply that the extra-dimensions are
 too tiny to see them. In fact, in order to fit our real world,  
 the sizes of the internal dimensions
 are constrained to be the Planck length.
 This beautiful answer has been considered to
 be a unique one for a long time. However, the recent discovery of  
 Dp-brane solutions has changed the situation. The Dp-brane is a kind of 
 soliton solution with the p-dimensional spatial and 1-dimensional
 temporal extension on which the boundary of open strings can be attached.
 While closed strings can freely propagate in the whole 10-dimensional
 spacetime. Since the open string contains the standard matter and the
 closed string contains the graviton, the D-brane solution yields 
 the idea of the braneworld where the standard matters are confined 
 to the 4-dimensional  braneworld, while the gravitons can propagate 
 in the 10-dimensional bulk spacetime. Interestingly, in the case of the 
 braneworld, the constraint on the sizes of the internal dimensions is
 not so stringent compared with the case of Kaluza-Klein mechanism. Indeed,
 the accelerator experiments in the laboratory can not give the strong
 constraints because the standard matters are confined to the 4-dimensional
 braneworld. Moreover, the experiment of the Newtonian force is rather 
 poor at scales smaller than 0.1 mm. Hence, the sizes of the extra-dimensions
 could be 0.1 mm in this braneworld picture. 
 
 Because of the possibility of the large extra-dimensions in the braneworld
 models, the early universe must be studied with careful considerations 
 of the effects of the bulk geometry. In this sense, the extra-dimensions
 can be regarded as the new elements of the universe. 
 
 To be more precise, what we want to explore is the effects of the bulk 
 geometry on the generation and the evolution of the cosmological fluctuations.
 In the conventional 4-dimensional cosmological models, curvature fluctuations
 with the flat spectrum are generated from the quantum fluctuations 
 during the inflationary expansion. 
 Then, its amplitude is freezed till the wavelength crosses the horizon again.
 Now, this curvature fluctuations can be observed  through the cosmic 
 microwave background radiation. 
 There are various possible sources to alter this standard scenario 
 in the braneworld models. The most obvious possibility is the collision
 between different braneworlds. In related to this case, there may be the
 effect of the radion, namely the distance between two braneworlds. 
 Moreover, there are effects due to gravitational waves propagating 
 in the bulk,  this is often  called as Kaluza-Klein (KK) effects. 
 
 There are many braneworld models. As a representative, here we will
 concentrate on the simplest Randall-Sundrum two-brane model~\cite{RS1}.
 In this paper, using the gradient expansion method, we formulate
 the general scheme to investigate effects of the bulk geometry. 
 As an application, we propose the born-again
 braneworld scenario and discuss its resemblance to
 the pre-big-bang model~\cite{pbb}. We also make observational predictions on
 the spectrum of the primordial gravitational waves which turns out to
 be very blue.

\section{Inflation on braneworld}

 Let us start with  the effective Friedmann equation 
\begin{eqnarray}
   H^2 = {\kappa^4 \mu^2 \over 12}
                      - {3 \over \ell^2 }
         + {\kappa^2 \over \ell} \rho + \kappa^4 \rho^2 
                     + {{\cal C} \over \alpha^4} \ ,
\end{eqnarray}
where $H$, $\alpha$ and $\rho$ are, respectively, the Hubble parameter, 
the scale factor and the total energy density of each brane.
The Newton's constant can be identified as $8\pi G_N = \kappa^2 /\ell$. 
 Here, ${\cal C}$ is a constant of integration associated with the mass of
a black hole in the bulk. 
 This constant ${\cal C}$ is referred to as the dark radiation
 in which the effect of the bulk is encoded.  
 The effective cosmological constant for the braneworld 
 becomes
\begin{equation}
    \Lambda_{\rm eff} = {\kappa^4 \mu^2 \over 12}
                      - {3 \over \ell^2 } \ ,
\end{equation}
where $\mu$ and $\ell$ are the tension of the brane and the curvature
scale in the bulk which is determined by the bulk vacuum energy, respectively.
 For $\kappa^2 \mu = 6/\ell$, we have Minkowski spacetime as a solution.
 In order to obtain the inflationary universe, we need the positive
 effective cosmological constant. In the brane world model, there are
 two possibilities. One is to increase the brane tension and the other
 is to increase $\ell$. The brane tension can be controled by the scalar
 field on the brane. The bulk curvature scale $\ell$ can be controled
  by the bulk scalar field. The former case is a natural extension of the
  4-dimensional inflationary scenario~\cite{maartens}. The latter possibility
  is a novel one peculiar to the brane model. 
   Recall that, in the superstring theory,  scalar fields  are 
 ubiquitous. Indeed, the dilaton and moduli exists in the bulk 
 generically, because they arise as the  modes associated with the 
 closed string. Moreover, when the supersymmetry is spontaneously broken,
 they may have the non-trivial potential. Hence, it is natural to consider 
 the inflationary scenario driven by these 
 fields~\cite{kobayashi}. 
 Now, we develop the formalism to investigate these inflating braneworlds.

\section{Gradient Expansion Method}

In this section, for simplicity,
we explain the gradient expansion method
 using the single-brane model.
We use the Gaussian normal coordinate system 
\begin{eqnarray}
ds^2 = dy^2 + g_{\mu\nu} (y,x) dx^\mu dx^\nu
\end{eqnarray}
to describe the geometry of the brane world. 
Note that the brane is located at $y=0$ in this coordinate system. 
Decomposing the extrinsic curvature $K_{\mu\nu} = -g_{\mu\nu,y} /2$
into the traceless and the trace part
$
  K_{\mu\nu} 
    = \Sigma_{\mu\nu }+ {1\over 4} g_{\mu\nu} K  \ , 
$
we obtain the basic equations which hold in the bulk;
\begin{eqnarray}
&&\Sigma^\mu{}_{\nu ,y}-K\Sigma^\mu{}_{\nu} 
	=-\left[
	\overset{(4)}{R}{}^\mu{}_\nu 
	-{1\over 4} \delta^\mu_\nu \overset{(4)}{R} 
        \right]
        \label{munu-trc} \ ,     \\
&&{3\over 4}K^2-\Sigma^\alpha{}_{\beta}\Sigma^\beta{}_{\alpha} 
	=\left[~
	\overset{(4)}{R}
	~\right] 
	+{12\over\ell^2}
	\label{munu-trclss} \ ,  \\
&&\nabla_\lambda\Sigma_{\mu}{}^{\lambda}  
	-{3\over 4}\nabla_\mu K = 0
	\label{ymu} \ ,
\end{eqnarray}
where  $\nabla_\mu $ denotes the
 covariant derivative with respect to the metric $g_{\mu\nu}$ and
 $\overset{(4)}{R}{}^\mu{}_\nu$ is the curvature associated with it.
 One also have the junction condition 
\begin{equation}
\left[
	K^\mu{}_\nu -\delta^\mu_\nu K
	\right]
	\Big|_{y=0} 
	={\kappa^2\over 2}
	\left(
	-\mu \delta^\mu_\nu+T^\mu{}_\nu
	\right) \ ,
\end{equation}
where $T_{\mu\nu}$ is the energy momentum tensor of the matter.
Recall that we are considering the $Z_2$ symmetric spacetime. 

The problem now is separated into two parts. First, we will solve
 the bulk equations of motion with the Dirichlet boundary condition
 at the brane, $g_{\mu\nu} (y=0 ,x^\mu ) = h_{\mu\nu} (x^\mu ) $.
 After that, the junction condition will be imposed at the brane.
 As it is the condition for the induced metric $h_{\mu\nu}$, it
 is naturally interpreted as the effective equations of motion
 for  gravity on the brane.
 
  In this paper, we will consider the low energy
  regime in the sense that the energy density of matter, $\rho$, 
  on the brane is smaller than the brane tension, \i.e., 
  $\epsilon\equiv \rho /\mu \ll 1$. 
  In this regime, we can use the gradient expansion method where 
  the above ratio $\epsilon$ is the expansion parameter~\cite{kanno1}.
 
 At zeroth order,  we can neglect the curvature term.  
 Anisotropic 
 term must vanish in order to satisfy the junction condition. 
 Now, it is easy to solve the remaining equations. The result is
$
    \overset{(0)}{K} = {4\over \ell}    \ .
$
Using the definition of the extrinsic curvature, 
we get the zeroth order metric  as
\begin{equation}
 ds^2 = dy^2 +  a^2 (y) h_{\mu\nu}(x^\mu ) dx^\mu dx^\nu  \ , \quad
    a(y)  = e^{-2{y\over \ell}}    \ ,
\end{equation}
where  the tensor $h_{\mu\nu}$ is  the induced metric on the brane. 
 From the zeroth order junction condition, we obtain
 the well known relation $\kappa^2 \mu = 6/\ell$.
 Hereafter, we will assume that this relation holds exactly.

The iteration scheme consists in writing the metric $g_{\mu\nu}$
 as a sum of local tensors built out of the induced metric on the
 brane, the number of gradients increasing with the order. 
Hence, we will seek the metric as a perturbative series
\begin{eqnarray}
     g_{\mu\nu} (y,x^\mu ) =
  a^2 (y) \left[ h_{\mu\nu} (x^\mu)  + \overset{(1)}{g}{}_{\mu\nu} (y,x^\mu)
  + \overset{(2)}{g}{}_{\mu\nu} (y, x^\mu ) + \cdots  \right]  \ , 
\end{eqnarray}
where $a^2 (y) $ is extracted 
 and we put the Dirichlet boundary condition
$  
     \overset{(i)}{g}{}_{\mu\nu} (y=0 ,x^\mu ) =  0    \ ,
$  
so that $g_{\mu\nu} (y=0, x) =  h_{\mu\nu} (x)$ holds at the brane. 
Other quantities can be also expanded as
\begin{eqnarray}
K^\mu{}_{\nu}&=&{1\over\ell}
	\delta^{\mu}_{\nu}
        +\overset{(1)}{K}{}^{\mu}{}_{\nu}
	+\overset{(2)}{K}{}^{\mu}{}_{\nu}+\cdots  \nonumber\\
\Sigma^\mu{}_{\nu}
	&=&\qquad
	+\overset{(1)}{\Sigma}{}^{\mu}{}_{\nu}
	+\overset{(2)}{\Sigma}{}^{\mu}{}_{\nu} + \cdots          \ .
\end{eqnarray}

The next order solutions are obtained by taking into account the 
terms neglected at zeroth order. 
Substituting the zeroth order metric into $\overset{(4)}{R} (g)$, 
we obtain
\begin{equation}
\overset{(1)}{K} = {\ell\over 6a^2} R(h)
\label{1:trc} \ .
\end{equation}
Hereafter, we omit the argument of the curvature for simplicity. 
Simple integration of Eq.~(\ref{munu-trc}) also gives the traceless part
 of the extrinsic curvature as
\begin{equation}
\overset{(1)}{\Sigma}{}^{\mu}{}_{\nu}={\ell\over 2a^2}
	(R^\mu{}_{\nu}-{1\over 4}\delta^\mu_\nu R)  
	+{\chi^{\mu}{}_{\nu}(x)\over a^4}
	\label{1:trclss}  \ ,
\end{equation}
where the homogeneous solution satisfies the constraints 
$
\chi^{\mu}{}_{\mu}=0   \ , \  \chi^{\mu}{}_{\nu|\mu}=0 \ .
$
 This term  corresponds to  dark 
 radiation at this order.  
 
 At this order, the junction condition can be written as
\begin{eqnarray}
  \left[~\overset{(1)}{K}{}^{\mu}{}_{\nu} 
	-\delta^\mu_\nu\overset{(1)}{K}~\right] \Bigg|_{y=0}  
	={\ell\over 2}\left(
	R^\mu{}_{\nu}-{1\over 2}\delta^\mu_\nu R
	\right)
	+\chi^\mu{}_{\nu}
	={\kappa^2\over 2}T^\mu{}_{\nu}  \ .
\end{eqnarray}
Thus, we have obtained the effective equation on the brane.

\section{Effective Action for Two-brane System}

 We consider  the two-brane system in this section.
 Without matter on the branes, we have the relation
 between the induced metrics, 
$g^{\ominus\hbox{-}\rm brane}_{\mu\nu}
=e^{-2d/\ell}g^{\oplus\hbox{-}\rm brane}
\equiv \Omega^2 g^{\oplus\hbox{-}\rm brane} $
where $d$ is the distance between the two branes.
Here, $\oplus$ and $\ominus$ represent the positive and the negative
 tension branes, respectively. 
Although $\Omega$ is constant for vacuum branes, it becomes the
function of the 4-dimensional coordinates if we put the matter on the brane.

Adding the energy momentum tensor to each of the two branes,
and allowing deviations from the pure AdS$_5$ bulk, the
effective (non-local) Einstein equations on the branes at low
energies take the form~\cite{kanno2},
\begin{eqnarray}
G^\mu{}_{\nu} (h )
	&=&{\kappa^2\over\ell}\overset{\oplus}{T}{}^{\mu}{}_{\nu} 
	-{2\over \ell}\chi^\mu{}_{\nu} \,,
\label{A:einstein}
\\
G^\mu{}_{\nu}(f)
&=&-{\kappa^2\over\ell}\overset{\ominus}{T}{}^{\mu}{}_{\nu} 
	-{2\over\ell}{\chi^\mu{}_{\nu}\over\Omega^4} \ .
\label{B:einstein}
\end{eqnarray}
where $h_{\mu\nu}=g^{\oplus\hbox{-}{\rm brane}}_{\mu\nu}$, 
$f_{\mu\nu}=g^{\ominus\hbox{-}{\rm brane}}_{\mu\nu}=\Omega^2h_{\mu\nu}$
and the terms proportional to $\chi_{\mu\nu}$ are 5-dimensional
Weyl tensor contributions which describe
the non-local 5-dimensional effect.
Although Eqs.~(\ref{A:einstein}) and (\ref{B:einstein})
are non-local individually, with undetermined $\chi_{\mu\nu}$,
one can combine both equations to reduce them to local equations
for each brane. Since $\chi_{\mu\nu}$ appears only algebraically,
one can easily eliminate $\chi_{\mu\nu}$ from Eqs.~(\ref{A:einstein}) 
and (\ref{B:einstein}). 
Defining a new field $\Psi = 1-\Omega^2$,  we find 
\begin{eqnarray}
\hspace{-3mm}
G^\mu{}_{\nu}(h)&=&{\kappa^2 \over \ell \Psi } 
	\overset{\oplus}{T}{}^{\mu}{}_{\nu}
	+{\kappa^2 (1-\Psi )^2 \over \ell\Psi } 
	\overset{\ominus}{T}{}^{\mu}{}_{\nu} \nonumber \\
&&	+{1\over\Psi}\left(\Psi^{|\mu}{}_{|\nu} 
  	-\delta^\mu_\nu\Psi^{|\alpha}{}_{|\alpha}\right) 
  	+{3 \over 2 \Psi (1-\Psi )}\left( \Psi^{|\mu}\Psi_{|\nu}
  	- {1\over 2}\delta^\mu_\nu\Psi^{|\alpha}\Psi_{|\alpha} 
  	\right) \ ,
  	\label{A:STG1} 	
\end{eqnarray}
where $|$ denotes the covariant derivative with respect to the metric
$h_{\mu\nu}$. Since $\Omega$ (or equivalently $\Psi$) contains
the information of the distance between the two branes,
we call $\Omega$ (or $\Psi$) the radion.
We can also determine $\chi^{\mu}{}_{\nu}$ by eliminating $G^{\mu}{}_{\nu}$ 
from Eqs.~(\ref{A:einstein}) and (\ref{B:einstein}). Then,
\begin{eqnarray}
\chi^{\mu}{}_{\nu}&=&-{\kappa^2(1-\Psi)\over 2 \Psi} 
	\left( \overset{\oplus}{T}{}^{\mu}{}_{\nu} 
	+(1-\Psi)\overset{\ominus}{T}{}^{\mu}{}_{\nu}\right)  \nonumber\\
&&	-{\ell\over 2\Psi} \left[ \left(  \Psi^{|\mu}{}_{|\nu} 
	-\delta^\mu_\nu  \Psi^{|\alpha}{}_{|\alpha} \right) 
	+{3 \over 2(1 -\Psi )} \left( \Psi^{|\mu}  \Psi_{|\nu}
  	-{1\over 2} \delta^\mu_\nu  \Psi^{|\alpha} \Psi_{|\alpha} 
  	\right) \right]   \ .  
  	\label{A:chi}
\end{eqnarray}
 Note that the index of $\overset{\ominus}{T}{}^{\mu}{}_{\nu}$ is to be raised
or lowered by the induced metric on the $\ominus$-brane, $f_{\mu\nu}$.
The condition  $\chi^\mu{}_{\mu}=0$ yields 
\begin{eqnarray}
\Box\Psi&=&{\kappa^2\over 3\ell}(1-\Psi )
	\left\{ \overset{\oplus}{T} + (1-\Psi)\overset{\ominus}{T}  
	                     \right\} 
	-{1 \over 2 (1-\Psi )} \Psi^{|\mu}\Psi_{|\mu} \ .
  	\label{A:STG2}
\end{eqnarray}
The effective action for the $\oplus$-brane which gives 
Eqs.~(\ref{A:STG1}) and (\ref{A:STG2}) is
\begin{eqnarray}
S_{\oplus}&=&{\ell \over 2 \kappa^2} \int d^4 x \sqrt{-h} 
	\left[ \Psi R - {3 \over 2(1- \Psi )} 
     	\Psi^{|\alpha} \Psi_{|\alpha} \right] \nonumber\\
    && \!\!\!\!\!\!\!\!\!\!
    	+ \int d^4 x \sqrt{-h} {\cal L}^\oplus 
      	+ \int d^4 x \sqrt{-h} \left(1-\Psi \right)^2 {\cal L}^\ominus  
      	\ .  
      	\label{A:action} 
\end{eqnarray}

We now extend the analysis to the two-brane system with the bulk scalar field
 $\phi$ coupled to the brane tension $\sigma(\phi)$
 but not to the matter ${\cal L}_{\rm matter}$ on the brane. 
 The action reads
\begin{eqnarray}
S&=& {1 \over 2\kappa^2 }\int d^5 x \sqrt{-g}  {\cal R} 
      - \int d^5 x \left[ {1\over 2} g^{AB} \partial_A \phi \partial_B \phi
      + U(\phi)       \right] 	\nonumber \\
	&&
	-\sum_{i=\oplus ,\ominus}  \int d^4 x 
	\sqrt{-g^{i-{\rm brane}} } \mu_i (\phi)
        +\sum_{i=\oplus ,\ominus}
	\int d^4 x \sqrt{-g^{i-{\rm brane}}} {\cal L}_{\rm matter}^i
        \ , \label{action-5d}
\end{eqnarray}
where $\kappa^2$, ${\cal R}$ and $h_{\mu\nu}$ are the gravitational constant, 
the scalar curvature in 5-dimensions constructed from the metric $g_{AB}$
 and the induced metric on the brane, respectively.  
 We assume the potential $U(\phi)$ for the bulk scalar field takes the form
\begin{eqnarray}
U(\phi)&=&-\frac{6}{\kappa^2\ell^2}+V(\phi)\ ,
\label{potential}
\end{eqnarray}
where the first term is regarded as a 5-dimensional cosmological constant
and the second term is an arbitrary potential function.
The brane tension is also assumed to take the form
$
\mu_i (\phi)=\mu_{i0}+\tilde\mu_i (\phi)\ .
$
The constant part of the brane tension, $\mu_{i0}$ is tuned so that
the effective cosmological constant on the brane vanishes. 
The above setup realizes a flat braneworld after inflation ends and 
the field $\phi$ reaches the minimum of its potential. 

 Most interesting phenomena occur at low energy
 in the sense that the additional energy due to the bulk scalar field is small, 
 $\kappa^2\ell^2V(\phi)\ll 1$, and the curvature on the brane $R$ 
 is also small, $R\ell^2\ll 1$.  
 Here, again, we can use the gradient expansion method explained above.
The effective action can be deduced as~\cite{kanno1}
\begin{eqnarray}
S&=&{\ell\over 2\kappa^2}\int d^4x\sqrt{-h}\left[
	\Psi R
	-\frac{3}{2(1-\Psi)}\partial^\alpha \Psi \partial_\alpha \Psi 
	-\kappa^2\Psi\left(\partial^\alpha \eta \partial_\alpha \eta
	+2V_{\rm eff}\right)
	\right]  
	\nonumber\\
&&
	+\int d^4x\sqrt{-h}~
	\overset{\oplus}{\cal L}
	+\int d^4x\sqrt{-h}~(1-\Psi)^2
	\overset{\ominus}{\cal L}  \ ,
\end{eqnarray}
where the boundary field $\eta = \phi (y=0,x)$ is defined and 
the effective potential takes the form
\begin{eqnarray}
V_{\rm eff}=\frac{1}{\ell}\left[~
	\frac{1}{\Psi}  {d \tilde \mu_{\oplus} \over d\eta}
	+\frac{(1-\Psi)^2}{\Psi} {d \tilde\mu_{\ominus} \over d\eta}
	~\right]
	+\frac{2-\Psi}{2}V (\eta) \ .
\end{eqnarray}
 As this is a closed system, we can analyze a primordial spectrum
 to predict the  cosmic background fluctuation spectrum.
 
In the  Einstein frame $g_{\mu\nu}=\Psi h_{\mu\nu}$ 
with $\Psi = 1/\cosh^2 {\kappa\over \sqrt{6\ell}}\rho$, 
the above action becomes
\begin{eqnarray}
S&=&{\ell\over 2\kappa^2}\int d^4x\sqrt{-g} R
	-\int d^4x \sqrt{-g} \left[
	\frac{1}{2}\partial^\alpha \varphi \partial_\alpha \varphi 
	+\frac{1}{2} \partial^\alpha \eta \partial_\alpha \eta
	   + V^{\rm E}_{\rm eff} \right]  \nonumber\\
&&      +\int d^4x\sqrt{-g} \cosh^4{\kappa\over \sqrt{6\ell}}\varphi
	\ \overset{\oplus}{\cal L}
	+\int d^4x\sqrt{-g} \sinh^4{\kappa\over \sqrt{6\ell}} \varphi
	\ \overset{\ominus}{\cal L}  \ ,
\end{eqnarray}
where
\begin{eqnarray}
V^{\rm E}_{\rm eff} =  
	 {d \tilde\mu_{\oplus} \over d\eta}
	 \cosh^4 {\kappa\over \sqrt{6\ell} }\varphi
	+ {d \tilde\mu_{\ominus} \over d\eta}
	\sinh^4 {\kappa\over \sqrt{6\ell} }\varphi  
	+\ell \left( \cosh^2 {\kappa\over \sqrt{6\ell} } 
	\varphi -{1\over 2}
	\right) V (\eta ) \ .
\end{eqnarray}

\section{Cosmological Evolution }

In order to study the cosmological evolution of the background spacetime
 and the fluctuations, typically, 
 one needs to analyze the following action
\begin{eqnarray}
 S = {1\over 2\kappa^2 } \int d^4 x \sqrt{-g} R 
   + \int d^4 x \sqrt{-g} \left[ 
    -{1\over 2} \partial^\mu \varphi \partial_\mu \varphi 
    -{1\over 2} e^{2b(\varphi )} \partial^\mu \chi \partial_\mu \chi
    - V(\varphi , \chi ) \right]  \ ,
    \label{maction}
\end{eqnarray}
where $e^b$ is chosen as 1 for the bulk inflation model
and $\cosh^2 {\kappa\over \sqrt{6\ell} }\varphi$ 
for the brane inflation model. 
Taking the metric
\begin{eqnarray}
ds^2 = \alpha^2 (\tau) \left[ -d\tau^2 +\delta_{ij} dx^i dx^j \right] \ ,
\end{eqnarray}
we have the equations of motion for the background fields 
\begin{eqnarray}
  &&  \varphi^{\prime\prime} + 2 {\cal H} \varphi^\prime +a^2 V_{\varphi}
  = b_{\varphi} e^{2b} \chi^{\prime 2}  \\
  && \chi^{\prime\prime} 
  + 2\left( {\cal H} + b_\varphi \varphi^\prime \right) \chi^\prime
  + e^{-2b} a^2 V_\chi = 0 \\
  && {\cal H}^2 =  {\kappa^2 \over 3} \left[ 
  {1\over 2} \varphi^{\prime 2} +{1\over 2} e^{2b} \chi^{\prime 2}
  + a^2 V \right]  \\
  && {\cal H}^\prime - {\cal H}^2 = -{\kappa^2 \over 2}
  \left[ \varphi^{\prime 2} + e^{2b} \chi^{\prime 2} \right] \ ,
\end{eqnarray}
where ${\cal H}=\alpha' /\alpha $ and the prime represents the derivative 
with respect to the conformal time $\tau$. Here, 
 the suffix of the functions $b,V$ means the derivative.
 The trajectory of the classical solution is parametrized by
 the proper length $\sigma$ in the superspace $(\varphi ,\chi )$, 
 namely
\begin{eqnarray}
   \sigma^\prime = \cos \theta \varphi^\prime 
                    + \sin \theta e^b \chi^\prime \ .
\end{eqnarray}
Here, $\theta$ is defined by
\begin{eqnarray}
 \cos \theta = {\varphi^\prime \over \sqrt{\varphi^{\prime 2} 
                     +e^{2b} \chi^{\prime 2} } } \ , \quad
 \sin \theta = {e^b \chi^\prime \over \sqrt{\varphi^{\prime 2} 
                     +e^{2b} \chi^{\prime 2} }} \ .
\end{eqnarray}
In terms of  this variable, one obtain
\begin{eqnarray}
   \sigma^{\prime\prime} + 2{\cal H} \sigma^\prime + a^2 V_\sigma =0
\end{eqnarray}
where
$
  V_\sigma = \cos \theta V_\varphi + e^{-b} \sin \theta V_\chi \ .
$
This is nothing but the equations of motion for the single
 scalar field if $V_\sigma$ depends only on $\sigma$. 
 Thus, the perturbation of the field $\sigma$ must be adiabatic
 and  that of the orthogonal field $s$ must be entropy perturbation.  
 Note that the following formula is useful for later calculation:
\begin{eqnarray}
  \theta^\prime = -{a^2 V_s \over \sigma^\prime} 
                 -\sigma^\prime b_\varphi \sin \theta \ ,
\end{eqnarray}
where we have defined
$
 V_s = e^{-b} \cos \theta V_\varphi - \sin \theta V_\chi \ .
$

\section{Cosmological Perturbations}

Now we consider the cosmological perturbations.
 As the action must be written by the gauge invariant quantities,
  we calculate the action in a completely gauge fixed manner.
   It is convenient to start with the longitudinal gauge
\begin{eqnarray}
   ds^2 = a^2 (\eta ) \left[ -(1+ 2A) d\eta^2 
       + (1-2\psi ) \delta_{ij} dx^i dx^j \right] \ .
\end{eqnarray}
 We also consider the fluctuation of the scalar fields
\begin{eqnarray}
   \varphi \rightarrow \varphi + \delta \varphi  \ , \quad
   \chi   \rightarrow \chi + \delta \chi \ .
\end{eqnarray}
By the simple substitution of the metric and the scalar fields into 
the action (\ref{maction}), 
we obtain the second order action  
\begin{eqnarray}
  S &=& {1\over 2\kappa^2 }\int d^4 x a^2 \left[
  -6\psi^{\prime 2} -4 A_{|i} \psi^{|i} +2 \psi_{|i} \psi^{|i}
                     \right. \nonumber\\
&& \left. \qquad  +6 {\cal H} (A+\psi) (A^\prime - \psi^\prime ) 
  + 3 ({\cal H} - {\cal H}^2 ) (A+\psi )^2 \right]  \nonumber\\
 &&  +\int d^4x a^2 \left[ {1\over 2}\left(
 \delta \varphi^{\prime 2} -\delta\varphi_{|i} \delta\varphi^{|i} \right)
     +{1\over 2} e^{2b} \left(
 \delta \chi^{\prime 2} -\delta\chi_{|i} \delta\chi^{|i} \right) 
                              \right. \nonumber\\
&& \qquad \quad \left.
    -{a^2 \over 2}\left( V_{\varphi\varphi} \delta\varphi^2
 + 2 V_{\varphi\chi} \delta\varphi \delta \chi 
 + V_{\chi\chi} \delta\chi^2 \right)
     - \left(A+ 3\psi \right) \left( \varphi^\prime \delta\varphi^\prime 
 + e^{2b} \chi^\prime \delta\chi^\prime \right) \right. \nonumber\\
&& \qquad \qquad \left. 
         -\left( A-3\psi \right) a^2 \left( V_\varphi \delta\varphi 
                     + V_\chi \delta\chi \right) 
       + A\left( A + 3\psi \right) \left( \varphi^{\prime 2} 
            + e^{2b} \chi^{\prime 2} \right) \right. \nonumber\\
&& \qquad  \left.   +\left( {1\over 2} \varphi^{\prime 2} 
             +{1\over 2} e^{2b} \chi^{\prime 2}
   \right) \left( -{1\over 2} A^2 -3 A\psi +{3\over 2} \psi^2 \right)
   + 2 b_\varphi e^{2b} \delta \varphi \chi^\prime \delta \chi^\prime
    \right. \nonumber\\
&& \qquad \left. 
   -b_\varphi e^{2b} \chi^{\prime 2} \delta\varphi \left( A+3\psi \right)  
                          +{1\over 2} e^{2b} \chi^{\prime 2} 
   \left( b_{\varphi\varphi} + 2b_\varphi^2 \right) \delta\varphi^2 
 \right]     \ .
\end{eqnarray}
 To simplify the action, we introduce
 the adiabatic and the entropy perturbations defined by
\begin{eqnarray}
  \delta \sigma = \cos \theta \delta \varphi
                  + e^b \sin \theta \delta \chi \ , \quad
  \delta s      = e^b \cos \theta \delta \chi 
                 - \sin \theta \delta \varphi  \ , 
\end{eqnarray}
respectively. 
Using these variables, we can rewrite the action as
\begin{eqnarray}
 2\kappa^2 S &=& \int d^4 x a^2 \left[
  -6\psi^{\prime 2} -4 A_{|i} \psi^{|i} +2 \psi_{|i} \psi^{|i}
                      -12 {\cal H} A\psi^\prime
 -2 \left( {\cal H}^\prime + 2{\cal H}^2 \right) A^2 \right. \nonumber\\
 &&  \left. + \kappa^2 \delta \sigma^{\prime 2}+ \kappa^2 \delta s^{\prime 2} 
 +\kappa^2 \theta^{\prime 2} \delta s^2 
 +\kappa^2 \theta^{\prime 2} \delta \sigma^2 \right. \nonumber\\
&&\left. - \kappa^2 \left( \delta\sigma_{|i} \delta\sigma^{|i}
 +\delta s_{|i} \delta s^{|i} +a^2 V_{\sigma\sigma} \delta \sigma^2 
 + a^2 V_{ss} \delta s^2 \right) 
  -4\kappa^2 a^2 V_s A \delta s  \right. \nonumber\\
&&\left.
  +\left\{ \kappa^2 \sigma^{\prime 2} \left( 1+ 2\sin^2 \theta \right)
  \cos^2 \theta + \kappa^2 \sigma^{\prime 2} b_{\varphi\varphi}
   + 2\kappa^2 \sigma^\prime b_\varphi \cos^2 \theta \sin \theta 
  \right. \right. \nonumber\\
&&\left.  \left. \!\!\!\!
  -\kappa^2 b_\varphi \left( 1 + \sin^2 \theta \right) a^2
  \left( V_\sigma \cos \theta + V_s \sin \theta \right) \right\}
  \delta s^2 \right. \nonumber\\
&&\left.   -2\kappa^2 a^2 V_\sigma A \delta s 
  - 2\kappa^2 \sigma^\prime A \delta \sigma^\prime 
   +6 \kappa^2 \sigma^\prime \psi^\prime \delta\sigma 
  + 4\kappa^2 {a^2 V_s \over \sigma^\prime} \delta \sigma^\prime \delta s
  \right. \nonumber\\
 &&\left.
  + 12 \kappa^2 {\cal H} {a^2 V_s \over \sigma^\prime} \delta \sigma \delta s
  + 4\kappa^2 {a^2 V_\sigma V_s \over \sigma^{\prime 2} }
            \delta \sigma \delta s  
   - \kappa^2 \sigma^{\prime 2} b_\varphi^2 \sin^2 \theta \delta \sigma^2
   \right. \nonumber\\
&&\left.
   +\kappa^2 b_\varphi a^2 V_\sigma \sin^2 \theta \cos \theta \delta \sigma^2
                -\kappa^2 b_\varphi \sin^3 \theta a^2 V_s \delta \sigma^2 
  \right]  \ ,
\end{eqnarray}
where
\begin{eqnarray}
V_{\sigma\sigma} &=&  V_{\varphi} \cos^2 \theta 
              + e^{-b} V_{\varphi\chi} \sin 2\theta 
              + V_{\chi\chi} \sin^2 \theta   \ , \\
V_{ss} &=& V_{\varphi} \sin^2 \theta 
              - e^{-b} V_{\varphi\chi} \sin 2\theta 
              + V_{\chi\chi} \cos^2 \theta   \ .
\end{eqnarray}
However, 
  to simplify the action, it is necessary to  move on to
  the uniform adiabatic gauge.  
Hence, we perform the gauge transformation
\begin{eqnarray}
    A_L = A_U + B_U^\prime + {\cal H} B_U \ , \quad
    \psi_L = \psi_U - {\cal H} B_U \ , \quad
    \delta \sigma_L = \sigma^\prime B_U  \ .
\end{eqnarray}
Notice that the entropy field is gauge invariant.
Using the relation
\begin{eqnarray}
 V_\sigma^\prime = \theta^\prime V_s - b_\varphi \sigma^\prime
 \cos\theta \sin \theta e^{-b} V_\chi + \sigma^\prime V_{\sigma\sigma}
\end{eqnarray}
we get
\begin{eqnarray}
  2\kappa^2 S &=& \int d^4 x a^2 \left[
  -6\psi^{\prime 2} -12 {\cal H} A\psi^\prime 
  -2\left( {\cal H} + 2 {\cal H}^2 \right) A^2 \right. \nonumber\\
 && \left.
  -4\kappa^2 a^2 V_s A \delta s - 2 \psi_{|i} \left( 2A-\psi \right)^{|i}
  -4 \triangle B \left( \psi^\prime + {\cal H} A \right) \right. \nonumber\\
&&\left.  +\kappa^2 \delta s^{\prime 2} -\kappa^2 \delta s_{|i} \delta s^{|i}
  +\kappa^2 \theta^{\prime 2} \delta s^2 -\kappa^2 a^2 V_{ss} \delta s^2
            \right. \nonumber\\
&&\left.
  +\left\{ \kappa^2 \sigma^{\prime 2} \left( 1+ 2\sin^2 \theta \right)
  \cos^2 \theta + \kappa^2 \sigma^{\prime 2} b_{\varphi\varphi}
   + 2\kappa^2 \sigma^\prime b_\varphi \cos^2 \theta \sin \theta 
  \right. \right. \nonumber\\
&&\left.  \left. \!\!\!\!
  -\kappa^2 b_\varphi \left( 1 + \sin^2 \theta \right) a^2
  \left( V_\sigma \cos \theta + V_s \sin \theta \right) \right\}
  \delta s^2 \right]  \ ,
  \label{uniform}
\end{eqnarray}
where we omitted the suffix for simplicity.
Now, we can obtain the reduced action
 by eliminating the unphysical degrees of freedom. 
Taking the variation with respect to $B$, we have
\begin{eqnarray}
    A = - {\psi^\prime \over {\cal H}}  \ .
\end{eqnarray}
Substituting this into Eq.(\ref{uniform}), we obtain our main result:
\begin{eqnarray}
  2\kappa^2 S &=& \int d^4 x a^2 \left[
  2 \left( 1-  {{\cal H}^\prime \over {\cal H}^2 }\right)
  \left( \psi^{\prime 2} -\psi_{|i} \psi^{|i} \right) 
  +4\kappa^2 {a^2 V_s \over {\cal H} } \psi^\prime \delta s 
                           \right. \nonumber\\
&&\left.  +\kappa^2 \delta s^{\prime 2} -\kappa^2 \delta s_{|i} \delta s^{|i}
   -\kappa^2 {\cal M}_s^2 \delta s^2 \right] \ ,
\end{eqnarray}
where
\begin{eqnarray}
   {\cal M}_s^2 
  &=& a^2 V_{ss} - \theta^{\prime 2}   
 - \sigma^{\prime 2} \left( 1+ 2\sin^2 \theta \right) \cos^2 \theta 
                  - \sigma^{\prime 2} b_{\varphi\varphi}
                   \nonumber\\
&&  
  - 2 \sigma^\prime b_\varphi \cos^2 \theta \sin \theta 
 + b_\varphi \left( 1 + \sin^2 \theta \right) a^2
  \left( V_\sigma \cos \theta + V_s \sin \theta \right) \ .
\end{eqnarray} 
Defining the new variable
\begin{eqnarray}
   {\cal U} = z \psi     \ ,   
   \quad  z = {a\sigma^\prime \over {\cal H}} \ ,
\end{eqnarray} 
we finally obtain  the canonically normalized action 
\begin{eqnarray}
  S &=& \int d^4 x \left[ {1\over 2}
  \left( {\cal U}^{\prime 2} -{\cal U}_{|i} {\cal U}^{|i} 
             + {z^{\prime \prime} \over z} {\cal U}^2 \right) 
  +2 {a^2 V_s \over {\cal H} } 
  \left({{\cal U} \over z}\right)^\prime \delta s 
                          \right. \nonumber\\
&&\left.  + {1\over 2}a^2 \delta s^{\prime 2} 
           - {1\over 2} a^2 \delta s_{|i} \delta s^{|i}
   -{1\over 2} {\cal M}_s^2 a^2 \delta s^2 \right]  \ .
\end{eqnarray}
The above action clearly shows the non-trivial coupling between
 the adiabatic and the entropy perturbations.


\section{Born-again braneworld}

In this section, as an application, we present the born-again braneworld
scenario~\cite{kanno3}. 
For simplicity, here, we we do not consider the bulk scalar field.
When we consider only the vacuum energy on each brane,
the effective potential for the radion takes the form depicted in Fig.1.
The unstable extrema corresponds to the deSitter two-brane system.

\begin{figure}[h]
\centerline{\hbox{\includegraphics[width=0.5\textwidth]{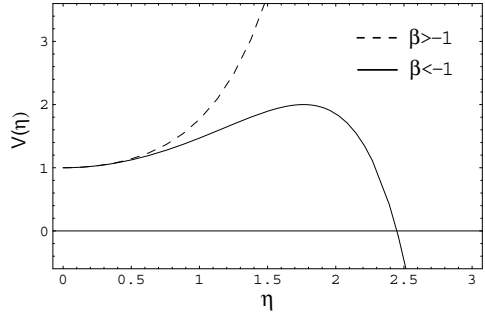}}}
\caption{Effective Potential for the radion.}
\end{figure}   

Our model has the features of both inflationary and pre-big-bang 
scenarios (see Fig.2). 
In the original frame, which we call the Jordan frame since
gravity on the brane is described by a scalar-tensor type
theory, the brane universe is assumed to be inflating due to an
  inflaton potential. 
 While, the radion, which represents the distance between the branes
and acts as a gravitational scalar on the branes,
 has non-trivial dynamics and theses vacuum branes can collide and 
    pass through smoothly.
After collision, it is found that 
the positive tension and the negative tension 
 branes exchange their role.  Then, they move away from each other,
and the radion becomes trivial after a sufficient lapse of time.
The gravity on the originally negative tension
brane (whose tension becomes positive after collision) will then
approach the conventional Einstein theory except for 
tiny Kaluza-Klein corrections.

\begin{figure}[h]
\centerline{\hbox{\includegraphics[width=0.5\textwidth]{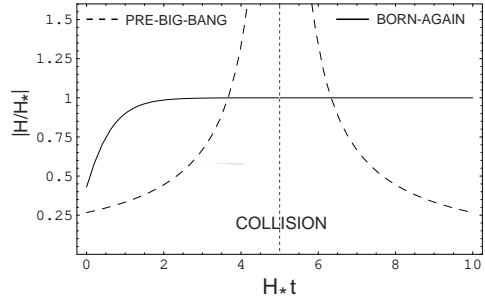}}}
\caption{The evolution of the Hubble constant in the Jordan frame.
 The de-Sitter spacetime is an attractor of the solution. 
 We also ploted the pre-big-bang solution in the Einstein frame.}
\end{figure}   

One can also consider the cosmological evolution of the branes in
the Einstein frame. Note that the two frames are indistinguishable
at present if our universe is on the positive tension brane after
collision. In the Einstein frame, the brane universe is contracting
before the collision and one encounters a singularity 
at the collision point. This resembles the pre-big-bang scenario.
Thus our scenario may be regarded as a non-singular realization
of the pre-big-bang scenario in the braneworld context.

\subsection{Gravitational waves}

 Now, we shall discuss the observational consequences of the scenario.
The universe  rapidly converges to the de-Sitter regime
 in the Jordan frame. As the metric couples with the radion,
 the non-trivial evolution of the radion field affects the tensor perturbations.
 This possibility descriminates our model from the usual inflationary 
 scenario.
 Let us consider the tensor perturbations in the Einstein frame:
\begin{equation}
ds^2 = \alpha^2 (\tau) \left[
       -d\tau^2 + (\delta_{ij} + h_{ij}) dx^i dx^j \right] 
\label{mtrc:tnsr-ptb}
\end{equation}
where $\alpha $ is the scale factor and 
$h_{ij}$ satisfy the transverse-traceless conditions, 
 $h_{ij}^{,j} = h^i_i =0$. 
As for the gravitational tensor perturbations, we have
\begin{equation}
    h_k^{\prime\prime} + 2{\cal H}h_k'+k^2h_k
    =0 \ ,
\label{tnsr:schrodinger-typ}
\end{equation}
where $h_k$ is the amplitude of $h_{ij}$ and ${\cal H}=\alpha'/\alpha$.
Since ${\cal H}\sim (2\tau)^{-1}$ near the collision point,
$h_k$ has  the spectral index $n=4$ for the gravitational waves 
 (with the spectral index defined by $P_h(k)\propto k^{n-4}$ as in
the case of scalar curvature perturbation; for the tensor perturbation,
the conventional definition is $n_T=n-1$).
 Provided that inflation ends right after collision,
 this gives a sufficiently
blue spectrum that can amplify the density parameter
 $\Omega_g$ by several order of 
magnitudes or more on small scales as compared to conventional
inflation models. Thus, there arises a possibility that
it may be detected by a space laser interferometer for
low frequency gravitational waves such as LISA.
  
\subsection{Inflaton perturbation}

 On the other hand, the inflaton does not couple directly
with the radion field. Hence, the inflaton fluctuations are expected
to give adiabatic fluctuations with a flat spectrum. 
 To calculate the inflaton perturbation rigorously,
one needs to introduce an inflaton field explicitly and
consider a system of equations fully coupled with
the radion and the metric perturbation. 
According to the estimation of the effect of the metric perturbation
induced by radion fluctuations on the inflaton perturbation, 
the inflaton fluctuations are not much affected 
by the radion fluctuations~\cite{kanno3}.
Thus, the inflaton fluctuations will have a standard
scale-invariant spectrum.

\section{Conclusion}
The general formalism for studying an inflationary scenario  
in the two-brane system is presented. 
 The gradient expansion method played a central role. 
 In particular, the second order action for the cosmological
 perturbations has been derived. This result should be
 useful when we discuss the quantization of the cosmological fluctuations. 
 As an application, we have proposed a born-again braneworld scenario. 
 As our braneworld is inflating and the inflaton has essentially
no coupling with the radion field, 
the adiabatic density perturbation with a flat spectrum is 
naturally realized.
  While, as the collision of branes mimics the pre-big-bang scenario~\cite{pbb},
the primordial background gravitational waves 
with a very blue spectrum may be produced.
This brings up the possibility that we may be able to see 
the collision epoch by a future gravitational wave
detector such as LISA.

\section*{Acknowledgements}
This work was supported in part by  Grant-in-Aid for  Scientific
Research Fund of the Ministry of Education, Science and Culture of Japan 
 No.14540258 (JS) and No. 155476 (SK) and also
  by a Grant-in-Aid for the 21st Century COE ``Center for
  Diversity and Universality in Physics".  

\end{document}